\documentclass[conference]{IEEEtran}
\IEEEoverridecommandlockouts
\usepackage{cite}
\usepackage{amsmath,amssymb,amsfonts}
\usepackage{algorithmic}
\usepackage{graphicx}
\usepackage{subcaption}  
\usepackage{enumitem}
\usepackage{textcomp}
\usepackage{xcolor}
\usepackage{authblk}
\usepackage[linesnumbered,ruled,vlined]{algorithm2e}
\usepackage{booktabs}  
\usepackage{threeparttable}  
\usepackage{float}
\usepackage{hyperref}
\usepackage{flushend}

\usepackage[all=normal,paragraphs=tight,floats=normal,mathspacing=normal,wordspacing=tight,charwidths=tight,mathdisplays=normal,leading=tight]{savetrees}

\def\BibTeX{{\rm B\kern-.05em{\sc i\kern-.025em b}\kern-.08em
    T\kern-.1667em\lower.7ex\hbox{E}\kern-.125emX}}
\begin{document}

\title{Physically constrained unfolded multi-dimensional \\ OMP for large MIMO systems
 \\
\thanks{This work is supported by the French national research agency (MoBAIWL project, grant ANR-23-CE25-0013 and RIS3, grant ANR-23-CMAS-0023)}
}

\author[1]{Nay Klaimi}
\author[2]{Clément Elvira}
\author[1]{Philippe Mary}
\author[1]{Luc Le Magoarou}

\affil[1]{Univ Rennes, INSA Rennes, CNRS, IETR-UMR 6164, Rennes, France}
\affil[2]{IETR UMR CNRS 6164, CentraleSupelec Rennes Campus, 35576 Cesson Sévigné, France}

\newcommand{\remNK}[1]{{\scriptsize \color{red} [Nay: #1]}}

\maketitle

\begin{abstract}
Sparse recovery methods are essential for channel estimation and localization in modern communication systems, but their reliability relies on accurate physical models, which are rarely perfectly known. Their computational complexity also grows rapidly with the dictionary dimensions in large MIMO systems. In this paper, we propose MOMPnet, a novel unfolded sparse recovery framework that addresses both the reliability and complexity challenges of traditional methods. By integrating deep unfolding with data-driven dictionary learning, MOMPnet mitigates hardware impairments while preserving interpretability. Instead of a single large dictionary, multiple smaller, independent dictionaries are employed, enabling a low-complexity multidimensional Orthogonal Matching Pursuit algorithm. The proposed unfolded network is evaluated on realistic channel data against multiple baselines, demonstrating its strong performance and potential.
\end{abstract}

\begin{IEEEkeywords}
MIMO, channel estimation, localization, Sparse recovery, Deep unfolding 
\end{IEEEkeywords}

\section{Introduction}
Multiple-input multiple-output (MIMO) systems play a crucial role in modern wireless communications due to their numerous advantages, such as enabling high data rates, achieving large bandwidth, and facilitating advanced beamforming \cite{Rusek2013,Larsson2014}. However, as the demand for higher data rates increases, wireless systems must utilize channels of increasingly high dimensions, which in turn makes tasks such as channel estimation and localization more challenging.
Channel estimation is of paramount importance for communication systems in order to optimize the data rate/energy consumption tradeoff. User localization is equally important to enable accurate positioning and optimize location-aware services and resource allocation.
To address these tasks, two complementary families of approaches have been widely explored. 

\begin{figure}[h] 
\vspace{-0.2cm}
    \centering
    \includegraphics[width=\columnwidth]{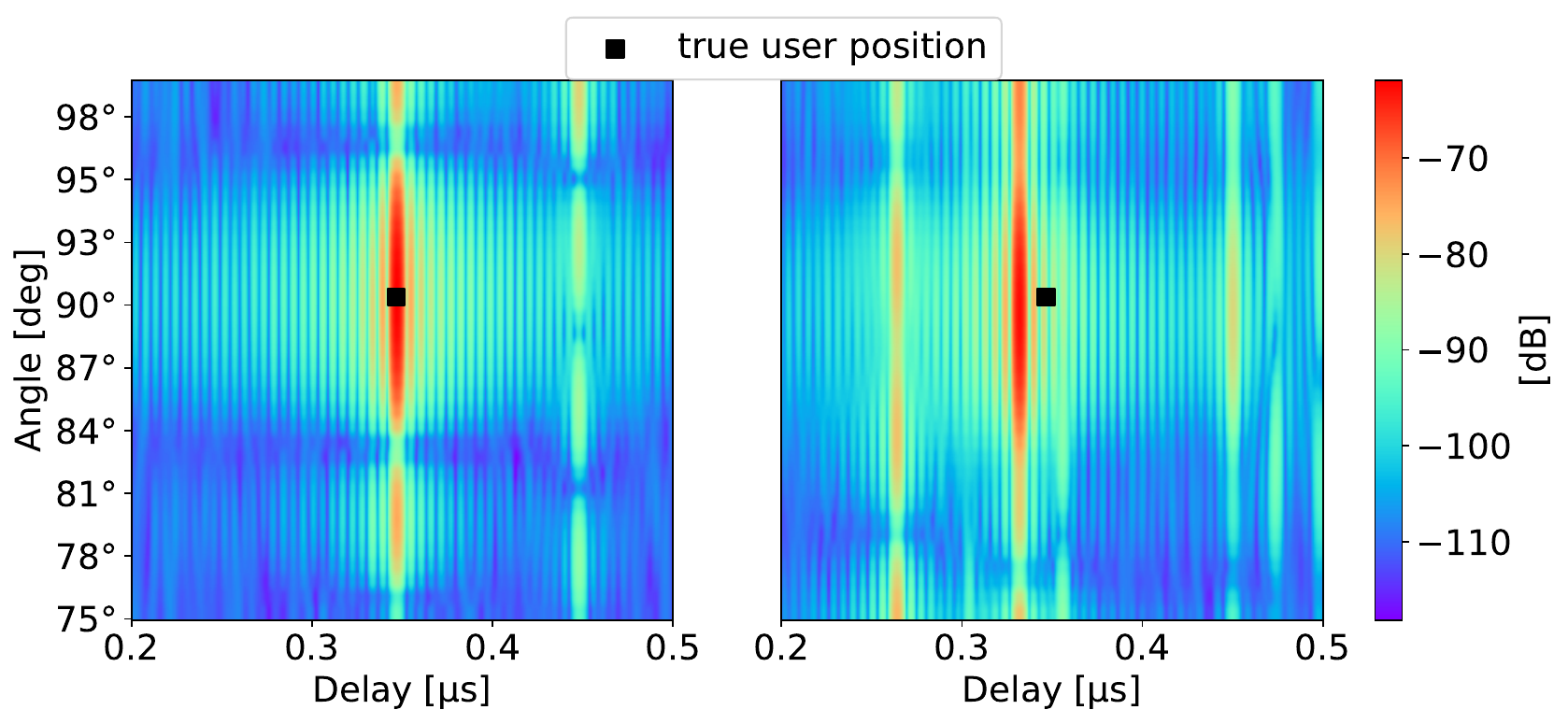} 
    \caption{Angle–delay maps: impairment-free scenario (left) and scenario with unaccounted hardware impairments (right).}
    \label{fig:ADM}
\vspace{-0.2cm}
\end{figure}

The first family comprises classical signal processing methods, namely model-based techniques that rely on prior knowledge of the system and channel characteristics. This family includes least-squares, matched filtering, and greedy algorithms for sparse recovery. While these methods are interpretable, mathematically tractable, and can offer performance guarantees under well-defined assumptions, their effectiveness degrades when the underlying system model is inaccurate, for example due to hardware impairments. Fig. \ref{fig:ADM} illustrates the effect of neglecting such impairments during estimation. The angle-delay map is computed by correlating the observed signal with steering vectors defined over a discretized angle-delay grid (see Section \ref{sec:syst_model}). At the first iteration of greedy algorithms for sparse recovery, this correlation is typically maximized. As shown in the figure, when impairments are present but impairment-free parameters are assumed, the algorithm may select spurious angle–delay pairs in the initial step, which subsequently leads to an inaccurate channel estimate.
The second family comprises machine learning (ML) methods, which are data-driven approaches that leverage large datasets to learn underlying patterns and relationships without requiring explicit system models \cite{O'shea2017,Wang2017}. These methods are particularly attractive in complex scenarios, where deriving an explicit mathematical description is difficult, as they rely on sampling and learning from large amounts of training data. However, ML methods often lack the interpretability and theoretical guarantees of classical techniques and can involve significant computational complexity.

A promising recent approach, model-based machine learning (MB-ML) \cite{shlezinger2023}, has emerged to address scenarios where the limitations of both model-based and ML methods become apparent. This approach leverages the prior knowledge to constrain ML methods, thus reducing their complexity. For instance, an iterative algorithm can be unfolded into a neural network, where each layer corresponds to one iteration of the algorithm, and the learnable parameters can be constrained by the physics of the system. This has been recently applied to MIMO systems for a variety of tasks, including channel estimation \cite{yassine2022}, hybrid precoding \cite{Lavi2023,Klaimi2025}, and direction-of-arrival estimation \cite{chatelier2024}.
\\
\noindent \textbf{Contributions.} This paper presents an enhanced MB-ML approach for channel estimation and localization in high-dimensional MIMO systems taking hardware impairments into account. Compared to existing approaches, the proposed training strategy achieves superior performance with greater efficiency. Moreover, it considers only physical learnable parameters, ensuring high interpretability and generalization. More specifically, the main contributions are as follows:
\begin{itemize}[leftmargin=*, labelsep=0.5em] 
    \item A deep-unfolded neural network architecture that performs channel estimation and localization, in high-dimensional MU-OFDM MIMO systems comprising an antenna array at the base station (BS) and multiple mobile stations (MS), each equipped with its own antenna array, operating over a large number of subcarriers. The proposed solution is highly performant and cost-effective.
    
    \item Within the proposed network, a generic way of handling hardware impairments that incorporates mutual coupling, antenna position errors within the arrays, antenna gain errors at both the BS and MS, as well as subcarrier frequency-offset errors.
    
    \item An extensive set of experiments on realistic synthetic channels demonstrating the effectiveness of the proposed method compared to prior art and illustrating its applicability in practical scenarios.
\end{itemize}

\noindent \textbf{Related work.} This work builds upon previous contributions \cite{yassine2022,Chatelier2023,Klaimi2025}, which also leverage deep unfolding for sparse recovery. While these approaches demonstrate strong performance, their computational cost grows significantly with system dimensionality. In contrast, the proposed method is designed to efficiently operate in high-dimensional settings. Moreover, whereas prior works rely on the Matching Pursuit (MP) algorithm \cite{Mallat1993}, this paper adopts the Orthogonal Matching Pursuit (OMP) algorithm \cite{Pati1993}, an evolution of MP yielding better performance.

This work is also related to the multidimensional Orthogonal Matching Pursuit (MOMP) algorithm \cite{Palacios2022,Bayraktar2024}, an approximate version of OMP designed for Kronecker-structured dictionaries, which effectively handles high-dimensional models. Since model perturbations in the dictionaries degrade the estimation in such algorithms(see Fig.~\ref{fig:ADM}), the Method of Optimal Directions (MOD) is employed in \cite{Bayraktar2024} to mitigate these effects. Although effective to some extent, MOD has several limitations: only one dictionary can be learned at a time without destroying the underlying Kronecker structure, and the resulting learned dictionary lacks a clear physical interpretation. In this work, we adopt a MB-ML framework via deep unfolding to jointly learn all impairment-related dictionaries, while preserving both their physical meaning and Kronecker structure. This approach enables improved performance, greater flexibility, and stronger interpretability.
\section{System model}
\label{sec:syst_model}
The considered system operates in the millimeter-wave (mmWave) band and adopts a multi-user MIMO architecture. It consists of a BS equipped with a uniform linear antenna array (ULA) of $N_B$ antennas, serving $M$ multi-antenna users, referred to as the MSs. Each user is also equipped with a ULA comprising $N_M$ antennas. Furthermore, the system operates in a multicarrier setting with $N_S$ subcarriers. 

\subsection{Physical channel model}
\label{ssec:phymod}

Let us consider a multipath channel and denote $K$ the number of propagation paths. Let $\mathbf{g}_B \in \mathbb{C}^{N_B}$ denote the complex gain vector of the BS antennas, and let $\{\mathbf{p}_i\}^{N_B}_{i=1}$ represent their positions with respect to the array centroid, where $\mathbf{p}_i \in \mathbb{R}^3$ corresponds to the $x,y,z$ coordinates. The operating wavelength is given by $\lambda = \tfrac{c}{f}$, where $c$ is the speed of light and $f$ is the carrier frequency. As described in \cite{yassine2022}, under the plane-wave assumption and assuming omnidirectional antennas (isotropic radiation patterns), the steering vector (SV) at the BS corresponding to the $k$-th path with angle of arrival (AoA) $\phi_k$ is
\[
\mathbf{e}_{s,B}(\phi_k) 
= \operatorname{diag}(\mathbf{g}_B)\big(\mathrm{e}^{-\mathrm{j} \tfrac{2\pi}{\lambda} \mathbf{p}^{\mathsf{T}}_1 \mathbf{u}(\phi_k)} ,\ldots, \mathrm{e}^{-\mathrm{j} \tfrac{2\pi}{\lambda} \mathbf{p}^{\mathsf{T}}_{N_B} \mathbf{u}(\phi_k)}\big)^\mathsf{T},
\]
Here, $\mathbf{u}(\phi_k) \in \mathbb{R}^{3}$ denotes the unit-norm vector corresponding to the azimuth angle $\phi_k$, and $\mathrm{diag}(\mathbf{g}_B)$ represents the diagonal matrix formed from the elements of $\mathbf{g}_B$.
An analogous expression holds for the MSs. The \(m\)-th MS, \(m=1,\dots,M\), has \(N_M\) antennas with complex gain vector \(\mathbf{g}_m \in \mathbb{C}^{N_M}\) and positions \(\{\mathbf{q}_{m,i}\}^{N_M}_{i=1}\) relative to the array centroid. Its steering vector for the \(k\)-th path with angle of departure (AoD) \(\psi_k\) is
\[
\resizebox{\columnwidth}{!}{$
\mathbf{e}_{s,m}(\psi_k)
= \operatorname{diag}(\mathbf{g}_m)
\big(\mathrm{e}^{-\mathrm{j} \frac{2\pi}{\lambda}
\mathbf{q}_{m,1}^\mathsf{T} \mathbf{u}(\psi_k)}, \dots,
\mathrm{e}^{-\mathrm{j} \frac{2\pi}{\lambda}
\mathbf{q}_{m,N_M}^\mathsf{T} \mathbf{u}(\psi_k)}
\big)^\top
$}
.\]
Let $\mathbf{f} = (f_1, \ldots, f_{N_S})^\mathsf{T} \in \mathbb{R}^{N_S}$ denote the vector of subcarrier frequencies. The frequency response vector (FRV) corresponding to the $k$-th path with propagation delay $\tau_k$ is
\[
\mathbf{e}_f(\tau_k) = \big(\mathrm{e}^{-\mathrm{j} 2 \pi f_1 \tau_k} ,\ldots, \mathrm{e}^{-\mathrm{j} 2 \pi f_{N_S} \tau_k}\big)^\mathsf{T} 
\in \mathbb{C}^{N_S},
\]
Under this model, the $m$-th user's channel $\mathbf{h}_m \in \mathbb{C}^{N_B N_M N_S}$ can be expressed in a vectorized form as
\[
\mathbf{h}_m = \sum_{k=1}^{K} \alpha_k \, \mathbf{e}_{s,B}(\phi_k) \otimes \mathbf{e}_{s,m}(\psi_k) \otimes \mathbf{e}_f(\tau_k),
\]
where $\alpha_k \in \mathbb{C}$ represents the complex gain of the $k$-th path. This linear combination is said to be \emph{sparse} when the number of paths $K$ is small compared to the channel dimensions. $\otimes$ denotes the Kronecker product; for $\mathbf{A} \in \mathbb{R}^{m \times n}$ and $\mathbf{B} \in \mathbb{R}^{p \times q}$, $\mathbf{A} \otimes \mathbf{B} \in \mathbb{R}^{mp \times nq}$ is the block matrix with $(i,j)$-th block of size $p \times q$ is given by $a_{ij}\mathbf{B}$; the definition extends to vectors viewed as column matrices.
Note that the channel can also be represented in tensor form as $\mathbf{H}_m \in \mathbb{C}^{N_B \times N_M \times N_S}$. Both representations are used interchangeably throughout this paper, depending on the context.

\subsection{Hardware impairments}
\label{impairments}
In practice, the SVs and FRVs are not perfectly known due to various hardware imperfections. These imperfections may arise from manufacturing tolerances, environmental variations, and mutual coupling effects between antennas at both the BS and the MS. Additionally, the FRVs can be affected by synchronization errors and local oscillator clock offset.

Therefore, any estimation strategy based on a physical model requires precise knowledge of system parameters (i.e., antenna positions and gains at both the BS and the MS, as well as frequency offsets). In reality, these true parameters are unknown. Our knowledge of the system is therefore \emph{nominal}, corresponding to an ideal impairment-free scenario.  
The nominal parameters of the BS are denoted by $\widetilde{\mathbf{p}}$ and $\widetilde{\mathbf{g}}_B$, while the true (unknown) parameters $\mathbf{p}$ and $\mathbf{g}_B$ include perturbations and are defined respectively as follows \cite{Chen2024}. For each antenna element $i \in \{1,\ldots,N_B\}$,

\[\mathbf{p}_i = \widetilde{\mathbf{p}}_i + \lambda \mathbf{n}_{p,i},\quad \mathbf{n}_{p,i} = \begin{pmatrix} 0 \\ \varepsilon_{p,i} \\ 0 \end{pmatrix} ,\]
for a ULA aligned with the $y$-axis, where $\varepsilon_{p,i} \sim \mathcal{U}(-\delta_p, \delta_p)$.
\[\mathbf{g}_{B,i} = a_i \mathrm{e}^{\mathrm{j}\varphi_{B,i}}, \quad
a_i = \widetilde{a}_i + n_{a,i}, \quad
\varphi_{B,i} = \widetilde{\varphi}_{B,i} + n_{\varphi,i},
\]
where $n_{a,i} \sim \mathcal{U}(-\delta_a,0)$ and
$n_{\varphi,i} \sim \mathcal{U}(-\delta_{\varphi_B}, \delta_{\varphi_B})$.

Analogously, for the $m$-th MS, the nominal parameters are denoted by
$\widetilde{\mathbf{q}}_{m}$ and $\widetilde{\mathbf{g}}_{m}$. For each antenna element $i \in \{1,\ldots,N_M\}$, \(
\mathbf{q}_{m,i} = \widetilde{\mathbf{q}}_{m,i} + \lambda \mathbf{n}_{q_{m,i}},
\quad
\mathbf{n}_{q_{m,i}} =
\begin{pmatrix}
0 \\ \varepsilon_{q_{m,i}} \\ 0
\end{pmatrix}\), where \(\varepsilon_{q_{m,i}} \sim \mathcal{U}(-\delta_q,\delta_q)\).

\noindent And \(\mathbf{g}_{m,i} = a_{m,i} \mathrm{e}^{\mathrm{j}\varphi_{m,i}},\quad
a_{m,i} = \widetilde{a}_{m,i} + n_{a_{m,i}}, \quad
\varphi_{m,i} = \widetilde{\varphi}_{m,i} + n_{\varphi_{m,i}}\), where \(n_{a_{m,i}} \sim \mathcal{U}(-\delta'_a,0)\) and \(n_{\varphi_{m,i}} \sim \mathcal{U}(-\delta_{\varphi_M},\delta_{\varphi_M}).\)

Let $\mathbf{C}_B \in \mathbb{C}^{N_B \times N_B}$ denote the mutual coupling matrix between the BS antenna elements. We adopt the coupling model from \cite{Ye2008}, assuming that each antenna is only coupled with its adjacent elements: 
\[\mathbf{C}_B = \operatorname{tridiag}(c_1,1,c_1),\]
where \emph{tridiag} denotes a tridiagonal matrix with main diagonal entries equal to $1$ and sub-/super-diagonal entries equal to $c_1$, the mutual coupling coefficient between adjacent antennas satisfying $0 \le |c_1| < 1$. In the \emph{nominal} impairment-free scenario, the coupling is neglected so the coupling matrix reduces to the identity: 
$\widetilde{\mathbf{C}}_B = \mathbf{I} \in \mathbb{R}^{N_B \times N_B}$.

For the subcarrier frequencies, it has been shown in \cite{DongKyu1998,Speth1999} that the frequency offset depends on the subcarrier index $i$, the oscillator parts-per-million (ppm) value $\xi$, and the subcarrier spacing $\Delta f$. This allows us to define nominal subcarrier frequencies $\widetilde{\mathbf{f}}$, with the true frequencies given by:
\(
f_i = \widetilde{f}_i + i \, \delta{f}= \widetilde{f}_i + i \, \xi\Delta{f}, 
\quad i = 1, \ldots, N_S.
\)
The corresponding \emph{nominal} SVs and FRV are then
\(
\widetilde{\mathbf{e}}_{s,B}(\phi_k), \quad
\widetilde{\mathbf{e}}_{s,m}(\psi_k), \quad
\widetilde{\mathbf{e}}_f(\tau_k),
\)
defined in the same way as their \emph{real} counterparts, but using the \emph{nominal} parameters.

\section{Problem formulation}

This paper addresses the joint uplink channel estimation and user localization problem. 
It is assumed that the BS receives pilot signals from the users. Each user transmits an orthogonal pilot sequence, allowing the BS to estimate the corresponding channel vectors. The number of pilot sequences is assumed sufficient such that, after matched filtering at the BS, observations of the form 
\begin{equation}
    \mathbf{Y} = \mathbf{H} + \mathbf{N} \in \mathbb{C}^{N_B \times N_M \times N_S},
\end{equation} 
are available \cite{Bayraktar2024}.
Here, $\mathbf{H}$ represents the previously defined channel $\mathbf{H}_{m,p}$ associated with the $m$-th MS at its $p$-th position; for simplicity, the indices $m$ and $p$ are omitted. The noise tensor $\mathbf{N} \sim \mathcal{CN}(\mathbf{0}, \sigma^2 \mathbf{I})$ represents additive complex Gaussian noise. The signal to noise ratio (SNR) can be computed as follows \[
\text{SNR} \triangleq 
\frac{\|\mathbf{H} \|_F^2}{N \, \sigma^2},
\]
where $N = N_B N_M N_S$ denotes the total number of elements in each channel tensor. In this context, channel estimation amounts to the design of an estimator $\hat{\mathbf{H}}(\cdot)$, via the following optimization problem,
\begin{equation} \label{eq:estim}
\underset{\hat{\mathbf{H}}(\cdot)}{\text{minimize}} \;\;
\mathbb{E} 
\|\hat{\mathbf{H}}(\mathbf{Y}) - \mathbf{H}\|_F^2,
\end{equation}
where the expectation is taken over noise realizations.
Similarly, denoting $\mathbf{l}\in \mathbb{R}^3$ the true location of the MS, localization corresponds to the problem
\begin{equation} \label{eq:loc}
\underset{\hat{\mathbf{l}}(\cdot)}{\text{minimize}} \;\;
\mathbb{E} 
\|\hat{\mathbf{l}}(\mathbf{Y}) - \mathbf{l}\|_2^2.
\end{equation}
Note that $\mathbf{Y}$ is already an unbiased estimator of the channel, obtained via a least-squares (LS) estimation procedure, and will be referred to as the LS estimator in the sequel.
However, more accurate estimators can be obtained by incorporating physical knowledge into their design. Namely, it is possible to use the physical channel model of section \ref{ssec:phymod}.
Let $\boldsymbol{\theta}$ denote the vector of physical parameters (e.g., antenna gains and positions) that determine $\mathbf{H}$, and $\hat{\mathbf{H}}_{\boldsymbol{\theta}}(\cdot)$ denote a channel estimator depending on the physical channel model ($\hat{\mathbf{p}}_{\boldsymbol{\theta}}(\cdot)$ for a physical position estimator). Such an estimator is highly dependent on hardware impairments, and thus requires calibration of the physical parameters. In order to do so, a database of observations $\{\mathbf{Y}_i\}_{i=1}^{M P}$, corresponding to $M$ MSs at $P$ distinct positions each and corresponding to channel tensors $\{\mathbf{H}_i\}_{i=1}^{M P}$ is considered. The calibration task then corresponds to the following optimization problem:
\begin{equation} \label{eq:Cost_funct}
\underset{\boldsymbol{\theta}}{\text{minimize}} \;\;
\frac{1}{MP}\sum_{i=1}^{MP} 
\|\hat{\mathbf{H}}_{\boldsymbol{\theta}}(\mathbf{Y}_i) - \mathbf{Y}_i\|_F^2,
\end{equation}
where the precise structure of $\hat{\mathbf{H}}_{\boldsymbol{\theta}}(\cdot)$ and its dependence on the physical parameters are described in the next section. This optimization problem can be interpreted as looking for the physical parameters $\boldsymbol{\theta}$ that allow the physics-constrained estimator $\hat{\mathbf{H}}_{\boldsymbol{\theta}}(\cdot)$ to be as close as possible to the LS estimator $\mathbf{Y}$ (projecting the LS onto the set of physical estimators), based on the rationale that incorporating physical constraints will denoise the LS estimator.

\section{Proposed approach}
This paper proposes a deep-unfolded greedy sparse-recovery algorithm designed for multidimensional dictionaries. The unfolding yields a physically parameterized neural network that executes the algorithm in its forward pass while learning the dictionaries through backpropagation, using only a few trainable parameters. This approach enables efficient joint channel estimation and localization with preserved interpretability and strong generalization.
The methodological elements are presented in the following subsections.
\subsection{Sparse Channel Representations}
\label{ssec:SCR}
The channel model introduced in \ref{ssec:phymod} implies that the channel admits a sparse decomposition in the angular and delay domains. To exploit this property for channel estimation, a discretized sparse representation is adopted by sampling these domains with predefined resolutions.
This leads to the construction of sparsifying dictionaries, where $\mathbf{D}_B$ and $\mathbf{D}_M$ span the angular domains at the BS and MS, respectively, and $\mathbf{D}_S$ spans the delay domain. The corresponding resolutions determine the number of atoms. Accordingly, three dictionaries are defined:
\begin{itemize}[leftmargin=*, labelsep=0.5em]
    \item \textbf{BS dictionary:} $\mathbf{D}_B \in \mathbb{C}^{N_B \times A_B}$, a dictionary of BS steering vectors for $A_B$ potential AoAs uniformly distributed from $0$ to $\pi$ relative to the ULA axis, including antenna gains and mutual coupling:
    \[
    \mathbf{D}_B \triangleq \mathbf{C}_B \,  
        \Big[ \mathbf{e}_{s,B}(\phi_1) \; \cdots \; \mathbf{e}_{s,B}(\phi_{A_B}) \Big].
    \]

    \item \textbf{MSs dictionaries:} $\mathbf{D}_m \in \mathbb{C}^{N_M \times A_M}$, a dictionary of the $m$-th MS steering vectors for $A_M$ potential AoDs (uniformly distributed as in the BS case),with $m=1\, \dots\, M$:
    \[\mathbf{D}_m \triangleq \mathbf{C}_m \,  
        \Big[ \mathbf{e}_{s,m}(\psi_1) \; \cdots \; \mathbf{e}_{s,m}(\psi_{A_M}) \Big].
    \] 

    \item \textbf{Delay dictionary:} $\mathbf{D}_S \in \mathbb{C}^{N_S \times A_S}$, a dictionary of frequency responses for $A_S$ potential path delays:
    \[
        \mathbf{D}_S \triangleq \Big[ \mathbf{e}_f(\tau_1) \; \cdots \; \mathbf{e}_f(\tau_{A_S}) \Big]. 
    \]
\end{itemize}

To consistently account for hardware impairments, we also define the corresponding \emph{nominal} dictionaries, obtained by replacing the true parameters with their nominal counterparts:

\[
\widetilde{\mathbf{D}}_B \triangleq\,\mathrm{diag}(\widetilde{\mathbf{g}}_B) \Big[ \widetilde{\mathbf{e}}_{s,B}(\phi_1) \; \cdots \; \widetilde{\mathbf{e}}_{s,B}(\phi_{A_B}) \Big],
\]
\[\widetilde{\mathbf{D}}_m \triangleq\,\mathrm{diag}(\widetilde{\mathbf{g}}_m) 
    \Big[ \widetilde{\mathbf{e}}_{s,m}(\psi_1) \; \cdots \; \widetilde{\mathbf{e}}_{s,m}(\psi_{A_M}) \Big],\]
\[\widetilde{\mathbf{D}}_S \triangleq \Big[ \widetilde{\mathbf{e}}_f(\tau_1) \; \cdots \; \widetilde{\mathbf{e}}_f(\tau_{A_S}) \Big].\]  

\subsection{Multidimensional Orthogonal Matching Pursuit MOMP}
The MOMP algorithm \cite{Palacios2022} is an extension of the standard OMP \cite{Pati1993}, 
designed for dictionaries that can be expressed as a Kronecker product
\(
\mathbf{D} = \bigotimes_{d=1}^{N_D} \mathbf{D}_d,
\)
where $N_D$ is the number of independent dictionaries; in this work, $N_D = 3$. 
The algorithm exploits this Kronecker structure to perform an approximate atom search independently within each smaller dictionary 
$\mathbf{D}_d \in \mathbb{C}^{N_d \times A_d}$, 
instead of operating on the single large dictionary $\mathbf{D}$, thereby avoiding the costly matrix operations associated with full high-dimensional dictionary atom selection. 
Table~\ref{table:MOMP} details the atom search procedure and compares the complexity of OMP and MOMP in terms of the number of atoms of the constituent dictionaries.
\begin{table}[h] 
\centering
\caption{Comparison of 3D OMP and MOMP atom selection}
\label{table:MOMP}
\resizebox{\columnwidth}{!}{%
\begin{tabular}{@{}p{0.48\linewidth} p{0.48\linewidth}@{}}
\toprule
\multicolumn{1}{c}{\textbf{OMP (Full 3D Search)}} & 
\multicolumn{1}{c}{\textbf{MOMP (Sequential Search)}} \\
\multicolumn{2}{@{}p{\linewidth}@{}}{%
\textbf{Input:} Channel $\mathbf{H} \in \mathbb{C}^{N_1 \times N_2 \times N_3}$, dictionaries $\{\mathbf{D}_d\in \mathbb{C}^{N_d \times A_d} \}_{d=1}^3$ \newline
\textbf{Output:} best selected atom indices $i_1,i_2,i_3$} \\
\midrule
\textbf{Correlation \& Atom selection:} & \textbf{Step 1: Correlation along $\mathbf{D}_1$} \\
$\mathbf{C} = 
\mathbf{D}_3^{\mathsf{H}} \times_{3} 
(\mathbf{D}_2^{\mathsf{H}} \times_{2} 
(\mathbf{D}_1^{\mathsf{H}} \times_{1} \mathbf{H}))$\tnote{*} & 
$\mathbf{C}_1 = \mathbf{D}_1^{\mathsf{H}} \times_{1} \mathbf{H}$ \\
$[i_1,i_2,i_3] = \arg\max_{i,j,k} |\mathbf{C}(i,j,k)|^2$ & 
$i_1 = \arg\max_i \|\mathbf{C}_1(i,:,:)\|_F^2$ \\
& \textbf{Step 2: Correlation along $\mathbf{D}_2$}\\ 
&$\mathbf{C}_2 = \mathbf{D}_2^{\mathsf{H}} \, \mathbf{C}_1(i_1,:,:),$\\ 
&$i_2 = \arg\max_j \|\mathbf{C}_2(j,:)\|_2^2$ \\
& \textbf{Step 3: Correlation along $\mathbf{D}_3$}\\ 
&$\mathbf{C}_3 = \mathbf{D}_3^{\mathsf{H}} \, \mathbf{C}_2(i_2,:),$\\ 
&$i_3 = \arg\max_k |\mathbf{C}_3(k)|^2$ \\
& \textbf{Step 4: Refinement}\\
&Repeat for $N_{\text{iterations}}$: refine $i_d$ \\
\midrule
\textbf{Complexity:} $\mathcal{O}(A_1 A_2 A_3)$ & \textbf{Complexity:} $\mathcal{O}(A_1 + A_2 + A_3)$ \\
Exact but complex & Approximate but much faster \\
\bottomrule
\end{tabular}}
\begin{tablenotes}
\footnotesize
\item[*] The operator $\times_n$ denotes tensor–mode multiplication along the $n$-th mode.
\end{tablenotes}
\end{table}\\
Comparing the theoretical complexity of the methods with respect to the dictionary size shows that classical OMP is no longer feasible as the number of atoms grows.
The rest of the MOMP algorithm steps applied to channel estimation are described in Algorithm \ref{algo:MOMP}.

\begin{algorithm}[h]
\caption{MOMP \cite{Palacios2022} (high level overview)}
\label{algo:MOMP}
\KwIn{Channel observation $\mathbf{y}$, dictionaries $\{\mathbf{D}_d\}_{d=1}^3$}

Initialize $I \gets \emptyset$, $\mathbf{r} \gets \mathbf{y}$, $\mathbf{D}_I \gets [\,]$\;

\Repeat{stopping criterion}{
    Select $[i_1,i_2,i_3]$ using $\mathtt{MOMP}(\mathbf{r}, \mathbf{D}_1, \mathbf{D}_2, \mathbf{D}_3)$ tab.\ref{table:MOMP}\;
    
    $I \gets I \cup \{(i_1,i_2,i_3)\}$\;
    
    Append atom: $\mathbf{D}_{1[:,i_1]} \otimes \mathbf{D}_{2[:,i_2]} \otimes \mathbf{D}_{3[:,i_3]}$ to $\mathbf{D}_I$\;
    
    $\mathbf{x}^\star \gets \arg\min_{\mathbf{x}} \|\mathbf{y} - \mathbf{D}_I \mathbf{x}\|_2$\;
    
    Update the residual: $\mathbf{r} \gets \mathbf{y} - \mathbf{D}_I \mathbf{x}^\star$\;
}
\KwOut{Denoised channel $\hat{\mathbf{h}} \gets\mathbf{y}-\mathbf{r}$}
\end{algorithm}

\subsection{Deep Unfolding}
This step corresponds to the dictionary learning stage of the proposed method, which aims at recovering the hardware impairments represented within the dictionaries. The estimation strategy summarized in Algorithm~\ref{algo:MOMP} is unfolded in this step as a neural network as proposed in \cite{yassine2022}, referred to as MOMPnet, which takes the observation $\mathbf{Y}$ as input and outputs a channel estimate $\hat{\mathbf{H}}$. The network weights consist of the set of previously defined physical parameters $\boldsymbol{\theta}$, including the antenna positions at the BS and MSs, antenna gains, subcarrier frequencies, and related quantities. These weights are initialized using the \emph{nominal} parameter values, which represent the imperfect knowledge of the system. The parameters are subsequently optimized via mini-batch gradient descent using backpropagation in order to minimize the channel estimation error in~\eqref{eq:estim}. This optimization is performed in an unsupervised manner, as no access to the true channel dataset $\{\mathbf{H}_i\}_{i=1}^{MP}$ is assumed. For each observation originating from user $m$, the backpropagation process updates the BS-related and subcarrier parameters, as well as only the parameters associated with the $m$-th MS.
Let $\mathcal{B}$ denote the current mini-batch of size $B$. The expectation in the risk function is approximated by averaging over the mini-batch observations, yielding a mini-batched formulation of \eqref{eq:Cost_funct}:
\[
\mathcal{C}(\boldsymbol{\theta}) = \frac{1}{B} \sum\limits_{\mathbf{Y} \in \mathcal{B}}
\left\| \hat{\mathbf{H}}_{\boldsymbol{\theta}}(\mathbf{Y}) - \mathbf{Y} \right\|_F^2.
\]

\section{Experiments} 
{\noindent\bf Settings.} The proposed framework is evaluated using realistic synthetic channels generated with the Sionna ray-tracing simulator \cite{sionnaRT2025}, for the Paris Étoile scenario at a carrier frequency of $f = 28\,\text{GHz}$. The setup comprises a BS equipped with an impaired ULA of $N_B = 16$ antennas, and $10$ MSs equipped with ULAs of $N_M = 8$ antennas, each subject to its own hardware impairments. Each MS moves randomly within a $200\,\text{m}$ radius around the BS, and $100$ distinct random positions per MS are used as training data, yielding $1000$ training observations. The mini-batch size is set to $B = 100$.
The subcarrier spacing is set to $120\,\text{kHz}$. A pilot subcarrier is inserted every $12$ subcarriers, resulting in an effective pilot spacing of $\Delta f = 120 \times 12\,\text{kHz} = 1.44\,\text{MHz}$, and a total of $N_S = 128$ pilot subcarriers is considered. In the end, this configuration yields channels of dimension $N_B N_M N_S = 16384$. 
To reflect realistic operating conditions, the SNR varies across channel realizations, and scenarios with average SNR of 0, 5, and 15~dB are considered. 
The considered BS hardware impairments are gain errors, position errors, and mutual coupling. The gain errors are modeled as uniformly distributed with spread parameters $\delta_a = 0.4$ for amplitude and $\delta_{\psi_B} = 0.4$ for phase, while the antenna position errors follow a uniform distribution with spread $\delta_p = 0.24$. For mutual coupling, the coupling coefficient is set to $c_1 = 0.15 \mathrm{e}^{-j\frac{\pi}{6}}$. 
At the MSs, only antenna position errors are considered, modeled as uniformly distributed with spread $\delta_q = 0.24$, with an independent realization for each $m = 1, \dots, M$.
For the subcarriers, no impairments are included in this set of experiments, since subcarriers impairments have much less impact in practice. Nevertheless, it would be possible to account for them in a similar manner by following the same strategy used for the BS and MS.
The nominal system corresponds to an impairment-free system in which all antenna gains are equal to one, the antennas are uniformly spaced by half a wavelength in all the ULAs, and no antenna coupling is present.
Dictionaries are generated according to what is described in section~\ref{ssec:SCR}, with $A_B = 160$, $A_M = 80$ and $A_S = 1280$, with cosines of the angles equally spaced between $-1$ and $1$, and delays equally spaced between $0$ and $\frac{c}{\Delta f}$. Note that with such parameters, the dictionary to be used within previously proposed unfolded sparse recovery methods that do not exploit the Kronecker structure \cite{yassine2022} \cite{Klaimi2025} would be of size $16384 \times 16384000$, leading to a prohibitive computational cost. For this reason, an unfolded version of MOMP has to be used with such high-dimensional systems.
\begin{figure}[h] 
\vspace{-0.3cm}
    \centering
    \includegraphics[width=\columnwidth]{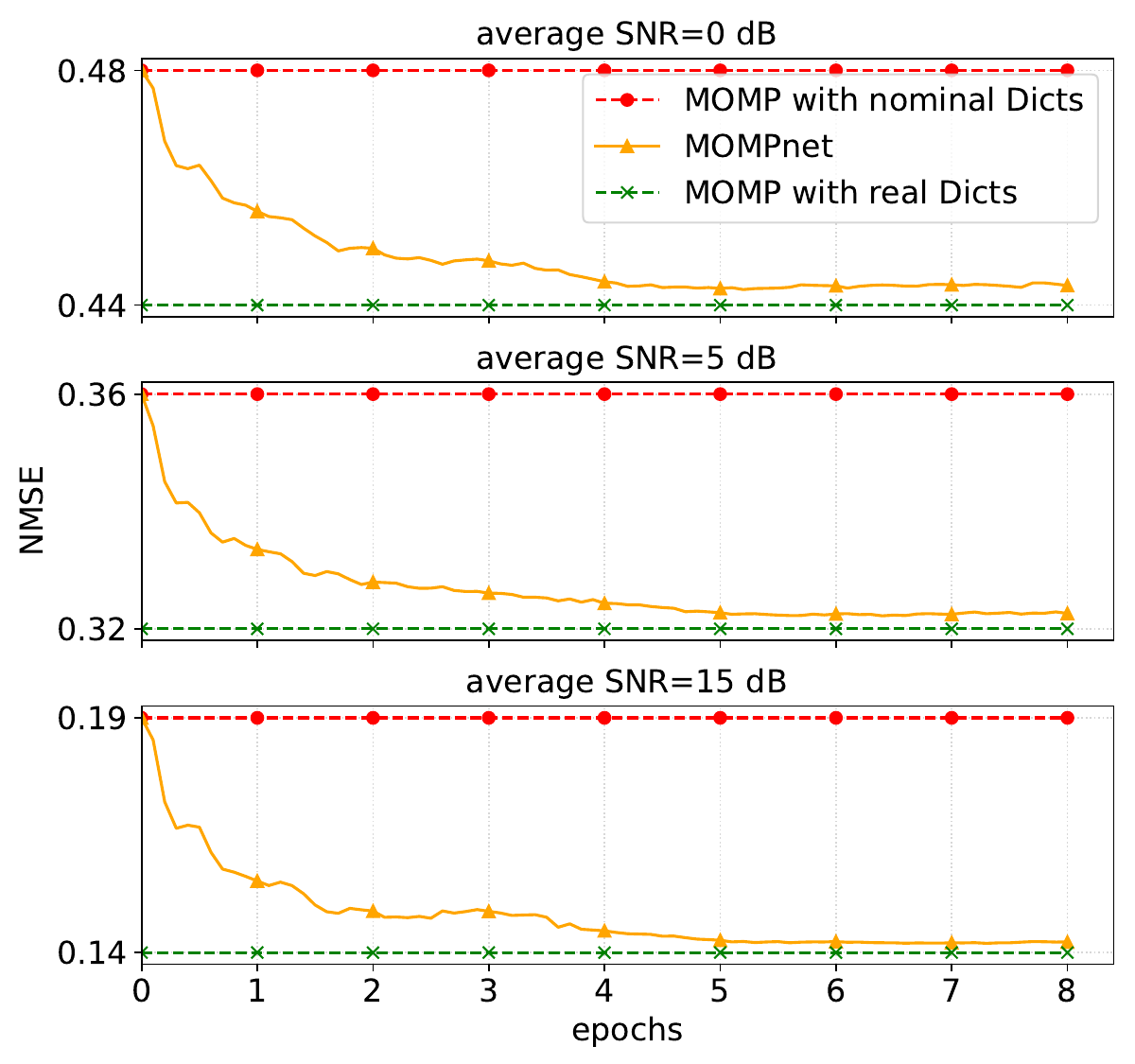} 
    \caption{Channel estimation performance on synthetic realistic channels for various SNRs}
    \label{fig:MOMPnet}
\vspace{-0.2cm}
\end{figure}\\
{\noindent\bf Channel estimation.} The results in Fig.~\ref{fig:MOMPnet} show the performance of the MOMPnet model for channel estimation. The performance metric is the normalized mean squared error (NMSE), averaged over a batch of $500$ channel realizations, unseen during training, defined as
$\text{NMSE} = \frac{\|\hat{\mathbf{h}} - \mathbf{h}\|_2^2}{\|\mathbf{h}\|_2^2}.$
The estimation error of the imperfect model (red) is shown to be effectively corrected by MOMPnet. After training on a relatively small number of observations, the performance of MOMPnet converges to that of the ideal model (green). Furthermore, this training is performed using a very small number of parameters, namely $130$ real parameters in the studied case containing $1$ BS and $10$ MSs.
\begin{figure}[h] 
    \centering
    \includegraphics[width=\columnwidth]{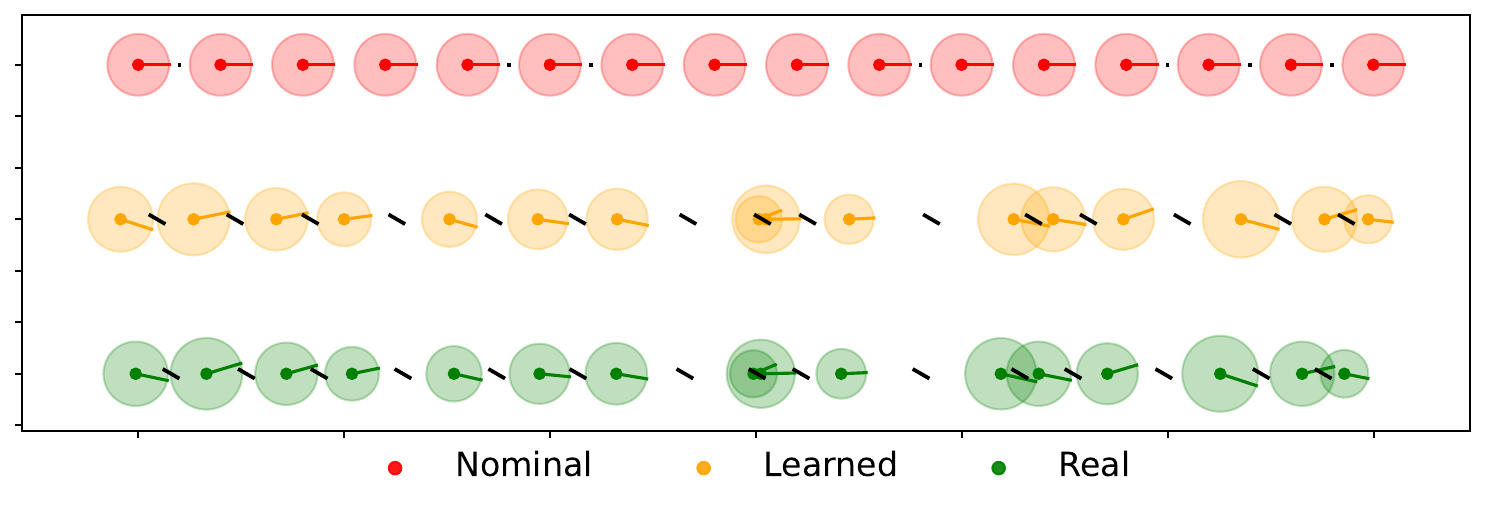} 
    \caption{ Learned parameters comparison at 5~dB SNR}
    \label{fig:learned_params}
\vspace{-0.2cm}
\end{figure}

\begin{figure}[h] 
\vspace{-0.2cm}
    \centering
    \includegraphics[width=\columnwidth]{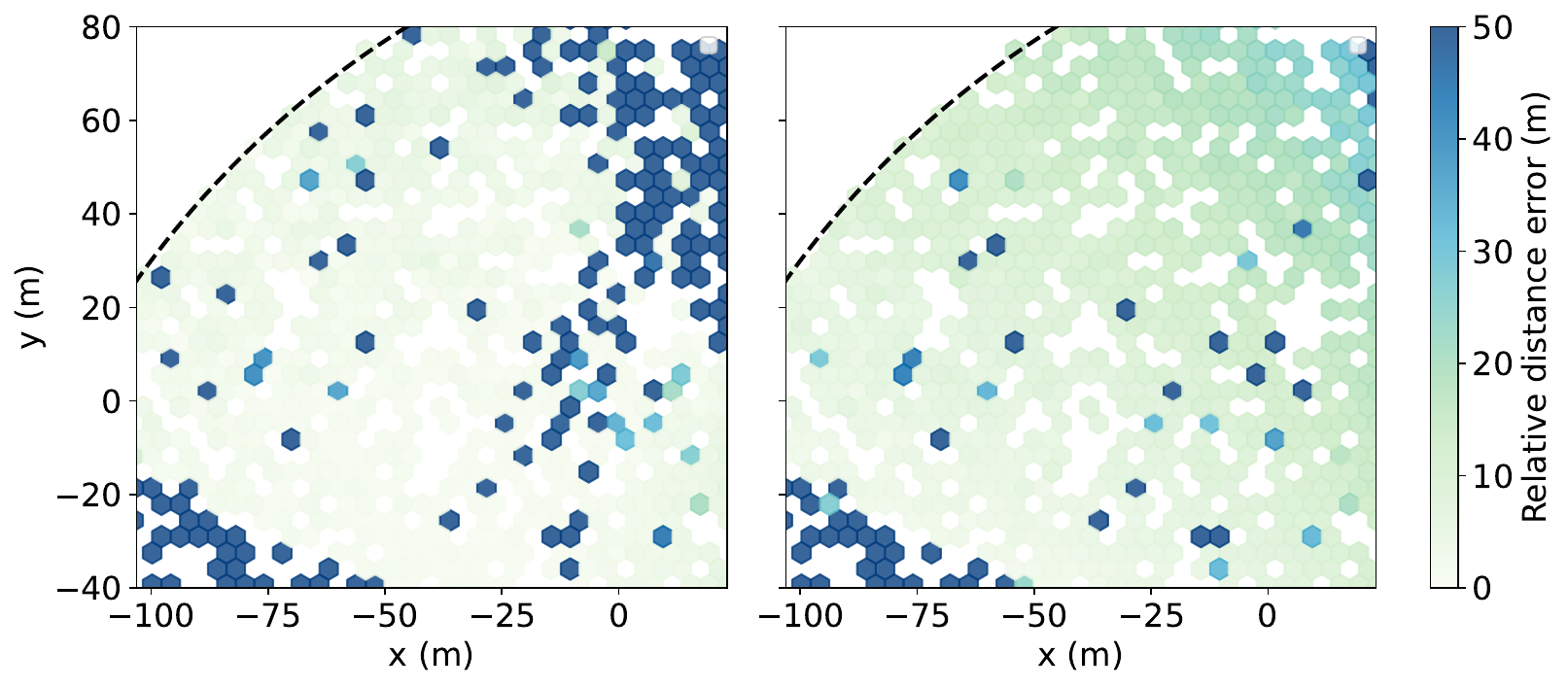} 
    \caption{Localization error at 5~dB before vs after training}
    \label{fig:localization}
\end{figure}

{\noindent\bf Interpretability and generalization.} This set of trained parameters has, as mentioned previously, a physical interpretation, as it represents the antenna parameters at the BS and the MS. Fig.~\ref{fig:learned_params} compares the learned parameters with the ideal and imperfect BS models. Each circle is centered at an antenna location, with its radius and angle representing the magnitude and phase of the complex gain, respectively. Black segments between antennas indicate the mutual coupling coefficient $c_1$, with length and orientation corresponding to its magnitude and phase. The imperfect (nominal) model parameters  are used as initialization for MOMPnet. After training, the model significantly reduces the mismatch with the ideal parameters. In particular, the mean absolute error, $\mathrm{MAE}=\frac{1}{N}\sum_{i=1}^{N}\lvert \hat{x}_i-x_i\rvert$, decreases from $0.06$ to $0.01$ for antenna gains and from $1.3\times10^{-3}$ to $7.3\times10^{-4}$ for antenna positions, while the learned coupling coefficient $c_1$ exactly matches the true value. This indicates that the unfolded network effectively captures the underlying antenna physics. This property provides high generalization and interpretability. For instance, training the model to improve channel estimation also enhances MS localization performance, as illustrated in Fig.~\ref{fig:localization}. MS positions are shown as points within the scene map delimited by the dotted black line, with point colors representing the localization error. After training, the average error is reduced by $9$~m. This improvement highlights the advantages of accurately learning antenna properties, which enables strong generalization beyond channel estimation, an ability difficult to achieve with task-specific classical neural networks.
\\
\begin{figure*}[t]
    \centering
    \begin{subfigure}{0.4\textwidth}
        \centering
        \includegraphics[width=\linewidth]{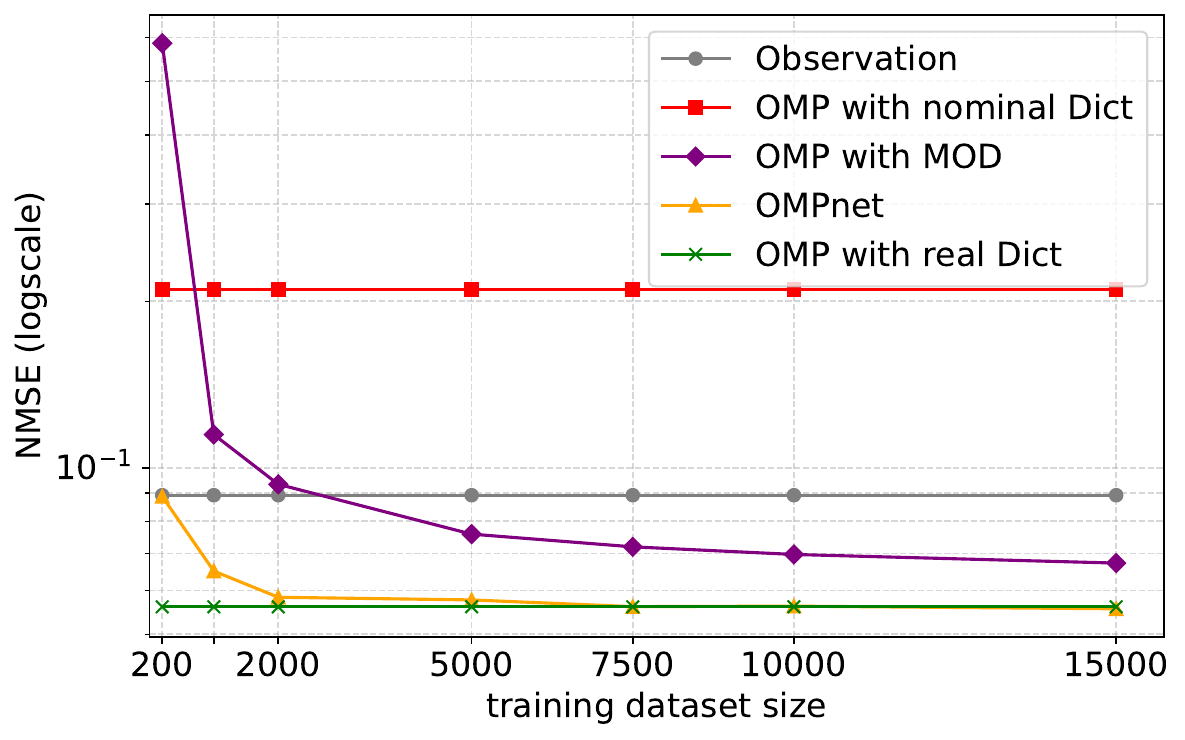}
        \captionsetup{labelformat=empty} 
        \caption{Fig.~\thefigure a NMSE vs. training dataset size at 10~dB SNR}
        \label{fig:MODobs}
    \end{subfigure}
    \hspace{0.2em}
    \begin{subfigure}{0.4\textwidth}
        \centering
        \includegraphics[width=\linewidth]{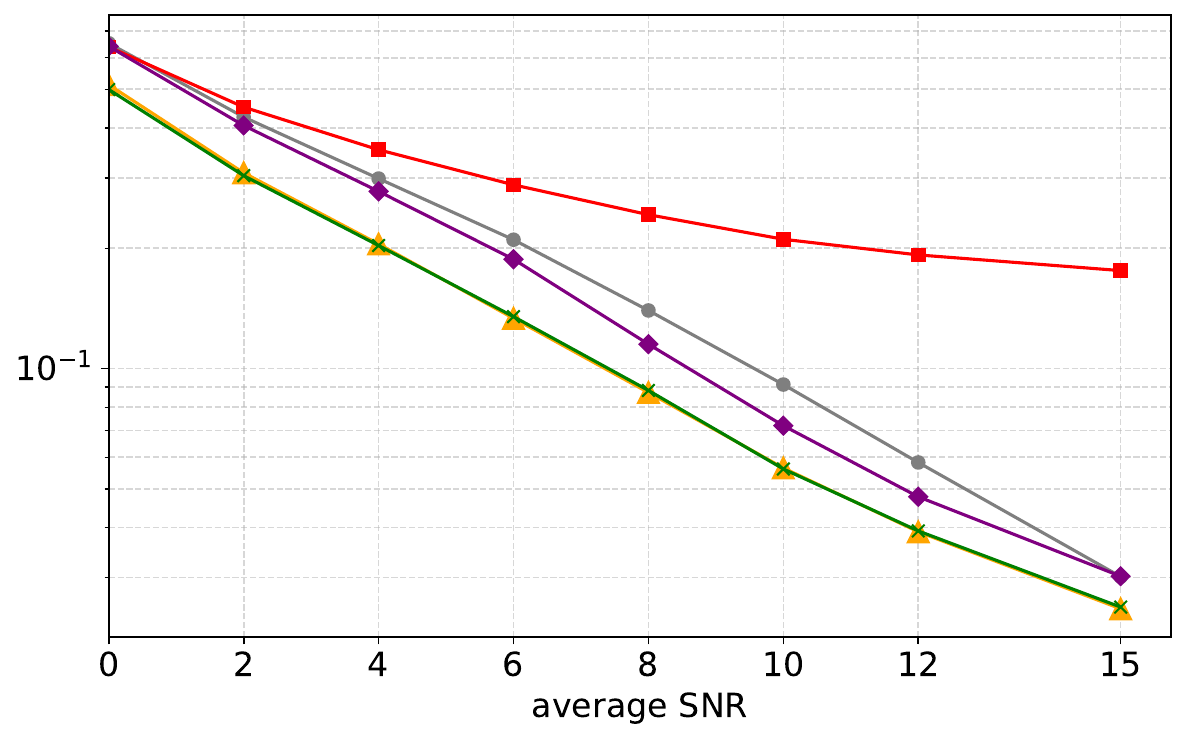}
        \captionsetup{labelformat=empty}
        \caption{Fig.~\thefigure b NMSE vs. SNR with 7,500 training samples}
        \label{fig:MODsnr}
    \end{subfigure}
\vspace{-0.4cm}
\end{figure*}
{\noindent\bf Comparison with MOD.} To enable a fair comparison with state of the art, the MOD algorithm, as used in \cite{Bayraktar2024}, is implemented. MOD requires to invert matrices whose size is the number of atoms and learns each entry of the dictionary as an independent parameter (without any physical structure). As mentioned previously, MOD is limited to training a single dictionary; therefore, it is compared with OMPnet, which corresponds to a reduced version of the proposed framework constrained to a single dimension. Figs.~\ref{fig:MODobs} and \ref{fig:MODsnr} compare the channel estimation error between the two methods as a function of the training dataset size and under different SNR conditions, respectively. In addition to its lower algorithmic complexity compared to the computationally intensive mathematical operations required by MOD, OMPnet is observed to achieve superior channel estimation performance, with faster convergence even when trained on a relatively small number of observations. Moreover, the learned dictionary in OMPnet, as in MOMPnet, retains a clear physical structure, as it is constructed from the learned physical parameters described earlier. In contrast, the dictionary learned by MOD does not preserve an explicit structure and lacks physical interpretability.
Furthermore, MOD requires a number of observations sufficiently larger than the number of columns in the dictionary in order to satisfy the method’s assumptions (for a matrix to be invertible), and all required observations must be collected prior to training, corresponding to an offline learning setting. By contrast, MOMPnet can be trained even with a single observation and can operate when observations are processed sequentially, corresponding to an incremental learning setting, as done for example in previous work on unfolding for channel estimation \cite{yassine2022,Klaimi2025}.

\section{Conclusion}
This paper introduces MOMPnet, a model-based neural network for high-dimensional channel estimation and localization. By representing the overall dictionary as a composition of lower-dimensional dictionaries, the method enables efficient dictionary learning even in very high dimensions. Leveraging deep unfolding, it can account for model inaccuracies due to hardware impairments, and the learning process can be carried out in an unsupervised, online manner, offering a novel and highly promising approach for large-scale systems.
The proposed method was also shown to outperform the state-of-the-art MOD algorithm. Although extensions of MOD to multidimensional settings have been reported in the literature~\cite{Roemer2014}, for the class of applications considered in this work, MOD remains computationally expensive and restrictive, without offering a performance advantage over MOMPnet.

\bibliographystyle{IEEEtran} 
\bibliography{references} 
\vspace{12pt}

\end{document}